\documentclass[letterpaper,11pt]{article}
\usepackage{amsfonts}
\usepackage[top=2.5cm, bottom=2.5cm, left=2.5cm, right=2.5cm]{geometry}

\usepackage{algorithm}
\usepackage{algorithmic}
\usepackage{graphicx}
\usepackage{caption}
\usepackage{abstract}
\usepackage{authblk}
\usepackage{tikz}
\usetikzlibrary{arrows,shapes}

\usepackage{subfigure}

\usepackage{setspace}
\renewcommand{\O}[1]{\mathcal{O}(#1)}

\title{Towards a Scalable Dynamic Spatial Database System}

\author[1]{Joaqu\'{\i}n Keller}
\author[1,2]{Raluca Diaconu}
\author[2]{Mathieu Valero}
\affil[1]{Orange Labs, Issy-les-Moulineaux, France 
}

\affil[2]{Laboratoire d'Informatique de Paris 6%
, 
Universit\'e P\&M Curie, 
 Paris, France

\vspace{5mm} \tt{\{joaquin.keller,raluca.diaconu\}@orange.com}\\
\tt{mathieu.valero@lip6.fr}

}

\date{}

\begin{document}
\maketitle


\begin{abstract}
With the rise of GPS-enabled smartphones and other similar mobile devices, massive amounts of location data are available. However, no scalable solutions for soft real-time spatial queries on large sets of moving objects have yet emerged. In this paper we explore and measure the limits of actual algorithms and implementations regarding different application scenarios. And finally we propose a novel distributed architecture to solve the scalability issues.

\vspace*{0.15cm}
\hspace*{-0.38cm}\textbf{Keywords:} Computational Geometry,  Distributed Computing, Spatial Databases, Moving Objects, Delaunay triangulations, R-Trees 
\end{abstract}

\section*{Introduction}

Today, hundreds of millions of smartphones and similar devices already generate a massive flow of location data. This huge amount of positions is poised to grow manifold, not only because smartphones are expected to be sold by billions but also because these devices, besides the now classical GPS, sport every year more sensors that can be used to acquire accurate positions. For example, wifi antennas help in collaboratively building detailed cartographies of wifi hotspots that in return are used to precisely locate any device by triangulation; accelerometers and gyroscopes ---reviving the old fashioned aircrafts' IMU\footnote{Inertial Measurement Unit}--- finely estimate the movements of smartphones allowing to compute their new positions using the dead reckoning technique.
Recently receiving academic and industrial attention, many imaginative location techniques have been proposed~\cite{WPNC2011}; they use all kinds of smartphone sensors to generate, for virtually any mobile device, accurate and frequent geolocation data.

Paradoxically, the massive amount of positions generated is rarely used for finding or searching mobile objects. Indeed, the typical use of a freshly produced location information is for querying data related with the location. Occasionally, this location information is used for tagging data to be queried in the preceding scheme. Sometimes this position information is stored in spatio-temporal databases \cite{Wolfson2004} for later mining. 

In the past centuries the opinion towards information has been {\em scientia potentia est}\footnote{knowledge is power}.
While gathering, representing and storing information are widely recognized as useful facilities, the ability to manipulate and process it becomes more and more important.
In particular answering queries related to the spatial properties of data under soft real-time constraints is an emerging requirement of several promising applications.


In this paper we argue that evanescent spatial data are underutilized mainly because of the poor performance of actual systems for soft real-time spatial queries on moving objects. We review the popular solutions based on search trees, and we propose the use of Delaunay triangulations for spatial indexes.
We define a general purpose benchmark from several application use cases, then analyze the poor performance in terms of scalability of existing solutions.
Our main contribution is to finally propose solutions to overcome these limitations by utilizing \emph{self adaptive zones}.

Firstly, we recall several application scenarios requiring soft real-time spatial queries.
Then, in Section~\ref{sec:dsd} we motivate our attention towards systems capable of handling a large number of moving objects and answering spatial queries about them in a timely manner. We call this kind of system Dynamic Spatial Database (DSD).
In Section~\ref{sec:benchmark} we define a benchmark inspired from the application scenarios and we investigate the limits for the two types of spatial indexes: R-trees and Delaunay triangulations.
Then we study the existing method to answer spatial queries, fixed zones, and other partial solutions to overcome scalability issues on particular cases.

Finally, we present a solution for scalable dynamic spatial databases ---a multi-level distributed architecture based on dynamic zones--- that supports an unlimited number of moving objects updating their position at arbitrary high frequencies.


\section{Application Scenarios for Dynamic Spatial Databases}
\label{sec:applications}
In order to identify relevant issues, we need to evaluate the limits of actual software solutions in application scenarios involving large numbers of moving objects. In particular, for some applications, meaningful time intervals are minutes while for other, events might occur every few milliseconds. Also, in some scenarios the number of moving objects does not exceed a few hundreds while in others hundred of millions are expected.

The application scenarios listed in this section extend previous lists \cite{Wolfson2004} but do not intend to be exhaustive. The idea is to exhibit enough engineering needs to outline the whole range of requirements for Dynamic Spatial Databases.

\subsection{Car Fleets}
To optimize resource allocation, goods, and people transportation, companies might want to know at any moment where their vehicles are. In the USA, the biggest of these companies manage  at best tens of thousands vehicles \cite{Grove2007} and could be satisfied with accuracies in the kilometer range. At full speed, for cars and trucks, it takes around 30 seconds to travel one kilometer. This is an acceptable mean time between two updates and we can take this value for our estimations.

We can easily imagine fleets composed of numbers of cars orders of magnitude above the actual fleets. The city of New York alone has more than 50000 taxis \cite{Schenkman2006} --including unlicensed for-hire vehicles.
And if the enthusiastic futurologists are right, in a near future self-driving cars will be the norm and the ``robotic taxis" would be counted in millions, far above the limits of actual solutions.

\subsection{Battlefield Awareness}
As the military concepts have moved toward network centric warfare \cite{Jameson2001}, it is now widely assumed that every vehicle and soldier will have a network connection and will precisely know its location. Also radars and similar sensors track or try to track every moving entity in the battlefield, friends and enemies alike.

The whole idea is to have an overall awareness of the situation to limit friendly fire and, more generally, to help the tactic organization at local or global levels.
So, as this is a matter of life and death, every object position should be known as timely and accurately as needed. Many military objects during combat tend to move --for offensive or defensive reasons-- as fast as possible, reaching up to supersonic speeds. This implies that position updates should occur every 10 to 1000 milliseconds, depending on the speed and the desired accuracy for the object.

Moreover, the introduction of unmanned vehicles leaves no limits to the number of moving objects evolving in a battlefield. If today a battle involves hundreds of thousands of mobile entities this number is poised to grow making it impossible to handle with today's dynamic spatial databases.


\subsection{Local Advertising}
Many predict that the future of advertising is mobile advertising, i.e., showing ads on mobile phones  \cite{Krumm2011}. Marketing strategy dictates to target advertising taking into account as many data as possible from the user, location being one of the most important. Most mobile marketing scenarios involves filtering out ad targets by location and not only by profile.

A typical scenario of location based advertising consists of sending ads to potential customers in the vicinity of a local commerce to induce impulsive buying behaviors. A example often cited is the restaurant with 20 meals left at 1 p.m.\  deciding to send 50\%-off coupons by text message to phone users nearby.

Let's take a conservative scenario where users' smartphones send their position every 10 minutes. Albeit marketers might be happier with 20 seconds update intervals, 10 minutes is still useful. 


\subsection{Location based Social Network, Geo Social Gaming, etc.}

With the rise of smartphone penetration, mobile location-based online social activities\cite{Scellato2011}, as networking, gaming or dating, have  recently attracted millions of users. One of the main motivations of the users of these services is the possibility to continue face to face the interaction started online. 

To make this happen it is necessary to find out who is nearby and can be reached in less than a given time. This time, depending of the nature of the interaction, it can be 20 seconds for some games or might be as long as 30 minutes for dating purposes. We can take these values --as position update intervals-- to estimate the maximum number of users that could be engaged simultaneously in one of these activities.


\subsection{Hybrid Reality, Virtual Worlds}
In virtual worlds, i.e.,  networked virtual environments, avatars are mobile objects and spatial queries or equivalent are issued to determine which avatars are relevant for a given user. This ensures a user gets all the information needed --she sees all the avatars in her visual field-- but, to avoid saturation, not much more. Furthermore, the minimum acceptable temporal resolution for vision is 10 to 30 frames per second. This gives us the position update rate for virtual worlds and hybrid reality.

In this area scalability is a well identified problem\cite{Horn2010,Gupta2009}, but assuming users enter a virtual world using a powerful well connected computer, most solutions rely on peer-to-peer networks \cite{Shun-YunHu2006,Nishide2008,Keller2002}.
However in the case of mixed reality \cite{Costanza2009} --\textit{a.k.a.} hybrid reality-- terminals are mostly lightweight computers running on battery and with a poor wireless connection. Power hungry peer-to-peer is hence not a valid solution and, as it is now common on applications for smartphones, computations should, when practical, be performed in the \textit{cloud}, i.e., on a datacenter.



\section{Spatial Databases}
\label{sec:dsd}
Spatial database management systems aim at supporting queries that involve the space characteristics of the underlying data. In a spatial database one can represent points, lines or polygons.
A spatial query selects objects based on geographic features, location or spatial relationship \cite{Manolopoulos2004}. It consists in searching within the available data the entries to satisfy a given geometric condition --e.g.,\ nearest neighbor, inclusion in a shape.

Spatial indexes are used to optimize spatial queries since regular indexes do not efficiently handle topology issues such as proximity or containment. Data structures commonly used for spatial indexes are \emph{R-trees} or \emph{Quadtrees}. Less common in spatial databases but frequently used in computational geometry are data structures based on \emph{Delaunay triangulations} --or \emph{Voronoi diagrams}. In this paper we study the performance of R-trees and Delaunay triangulations as spatial indexes for soft real-time spatial queries.

We raise here the problem of implementing a large highly dynamic spatial database --i.e.,\  with a large number of objects constantly moving and frequently updating their positions. We name a system capable of handling the trade-off between soft real-time performance necessary for updates or queries, and a proper data representation a \emph{dynamic spatial database(DSD)}.

\subsection{Dynamic spatial databases}

In many cases applications need the ability to represent moving entities that frequently change their positions. Applications might also want to query the entries using their current positions. Hence, the \emph{spatial} character of the applications.

In this setting the objects are constantly moving. The queries are meant to address the most recent configuration of the objects. Therefore the interest appears for updating the existing data rather than just storying it. Hence the \emph{dynamic} nature of the systems.

On the other hand, without additional attributes (properties) regarding the objects, there is no distinction between a set of objects and a collection of positions which, at any moment will be in some configuration, with no consistency with the previous ones. But preserving and manipulating information about objects is done by \emph{databases}.

For the remainder of this paper the notion of \emph{dynamic spatial databases} will denote this specific kind of systems described above. Contrariwise to the antithetical  appearance of the term, it emphasizes the trade-off between precision of the result and performance for these particular systems. If updates are applied to the database very often, the results are up-to-date, yet the update load becomes very high and unfeasible. Conversely, if updates are sent less frequently, the answers are outdated and erroneous.

A slightly different term, Moving Objects Databases (MOD) \cite{Wolfson2002a, Wolfson2004}, is used to denote Spatio-Temporal Databases.
These databases keep a history of all previous positions and are optimized to answer spatio-temporal queries (in the past only). Unlike MODs, DSDs only store (temporarily) the current position and focus on answering queries about the present configuration --i.e.\ the most recent configuration. The key feature is that DSDs must be able to answer real-time spatial queries. In addition DSDs do not exclude the possibility to store data for later manipulation, but this is not their main purpose. 

\section{Algorithms for 2D spatial queries}
\label{sec:benchmark}
A sphere is a good approximation of the Earth's surface. Since most human activities happen on this surface, the geography is often  represented in a two-dimensional Euclidean space, using latitude and longitude coordinates. For sure, this representation does not fully take into consideration rare cases where altitude might matter --when objects are inside skyscrapers, caves, or flying.
However it is possible to treat extra dimensions as additional attributes.
Moreover all aforementioned use cases use two-dimensional spaces.
Therefore, this paper covers the two-dimensional case exclusively.

We survey the fundamental methods to efficiently answer spatial queries on a large number of objects with today's dynamic spatial databases. The queries we focus on are insertions, deletions, displacements, and a spatial query. Two of the most common spatial queries are \emph{range query} and \emph{k nearest neighbors search (kNN)}) \cite{Zhang2003}.


The naive solution for range query (resp. kNN) problem is to compute the distance from the query object to each other object, keeping track of all the objects contained in the given range (resp. the best $k$ candidates) at any moment.
The time complexity of this naive approach is far from optimal. The worst case and the average  complexity for the query are $\O{n}$. 
This problem is overcame by using an adequate data representation that eliminates multiple candidates, thus lowering the costs of algorithms for spatial queries.

R-tree indexes are designed for accessing polygons and provide efficient algorithms for range search. The branch and bound algorithm for nearest neighbors search \cite{Roussopoulos1995} covers the tree as range query with additional operations at each step which will make kNN considerably less efficient than range query. On the other hand, Delaunay triangulations are best suited for proximity search and for answering range queries will have to perform at least a line walk to the queried area. Therefore, employing range search for R-trees and kNN for Delaunay triangulations will allow us to compare the performances of the two indexes.

\subsection{Experiments and settings}
\label{sec:experiments}

In order to examine the limits of the existing solutions we use a round based simulator.
To simulate a population of moving objects, each round is made of three steps:
\begin{enumerate}
	\item {\bf objects join the system:} each object is added to the spatial index.
	\item {\bf spatial queries are performed:} from a new position a spatial query (range query or nearest neighbor) is performed in the spatial index.
	\item {\bf objects move:} each object is removed from the spatial index then inserted with the new coordinates.
\end{enumerate}

In our experiments a simulation runs for 40 rounds. The number of objects added, queried, and displaced is the same for every round but, since objects are not removed from the index, at each round the size of the database will increase.
Initially, objects are placed uniformly at random on the geographical surface.
Objects' displacement is a teleportation: a fresh random position replaces the current one.

Furthermore, for an approach to the natural human distribution and mobility we have executed the tests on a Pareto distribution of objects where the movements are performed by L\`evy flights. However the benchmark witnesses the same behavior for each algorithm. We therefore present only the results for the uniform distribution.

The simulator is written in {\em Python}. 
The tests have been performed on a computer equipped with an Intel Core 2 Quad CPU Q9400  at 2.66 GHz, running a pre-installed Ubuntu 11.04 operating system. The system had 4 GB RAM and did not use disc caching during the tests. 
Each function was measured the processor time using \texttt{time.clock()}.

\subsection{R-trees}

{\em R-trees} \cite{Guttman1984}, provide access to spatially indexed multidimensional data. They are an extension of B-trees for a balanced representation of objects in space.

\begin{figure}[!ht]
	\centering
	\includegraphics[width=0.5\linewidth]{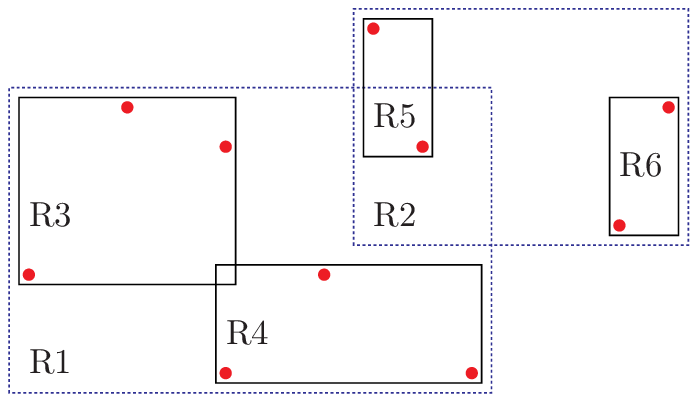}
	\caption{R-tree example}
	\label{fig:R-tree}
\end{figure}

The nodes are tuples $(r, oid)$ where $oid$ is an object identifier pointing to a data object, and $r = (x_{low}, x_{high}, y_{low}, y_{high})$ is a two-dimensional \emph{minimal bounding rectangle}, given by the coordinates of low left and up right corners. As for B-trees, the nodes form a balanced tree structure where the objects are stored in the leaves. The space required to store the full data structure is linear with respect to $n$. 

The depth of an R-tree storing $n$ objects is $\O{\log{n}}$.
The locate routine requires $\O{\log{n}}$ time. The insertion and deletion of a node, apart from the location procedure, must also re-balance the resulting tree. The complexity remains logarithmic.

Point displacement consists in removing and then inserting the object at the new coordinates. Finally, the algorithm for range query will use the bounding boxes to decide whether or not to search in a subtree. Therefore, most of the nodes will never be visited resulting in a logarithmic complexity.

\subsection{R-tree implementations}

\begin{figure}
\centering
	\subfigure[Insertion]{ 
		\includegraphics[width=0.4\linewidth]{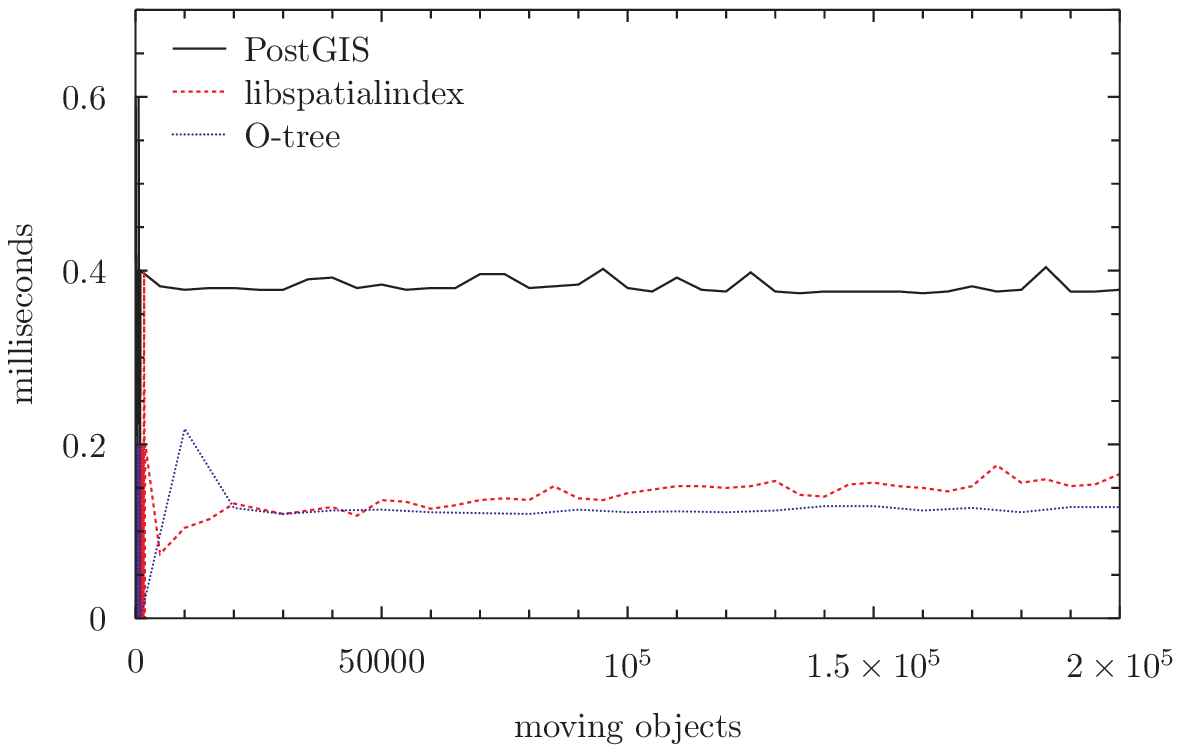}
	}
	\subfigure[Moving]{
		\includegraphics[width=0.4\linewidth]{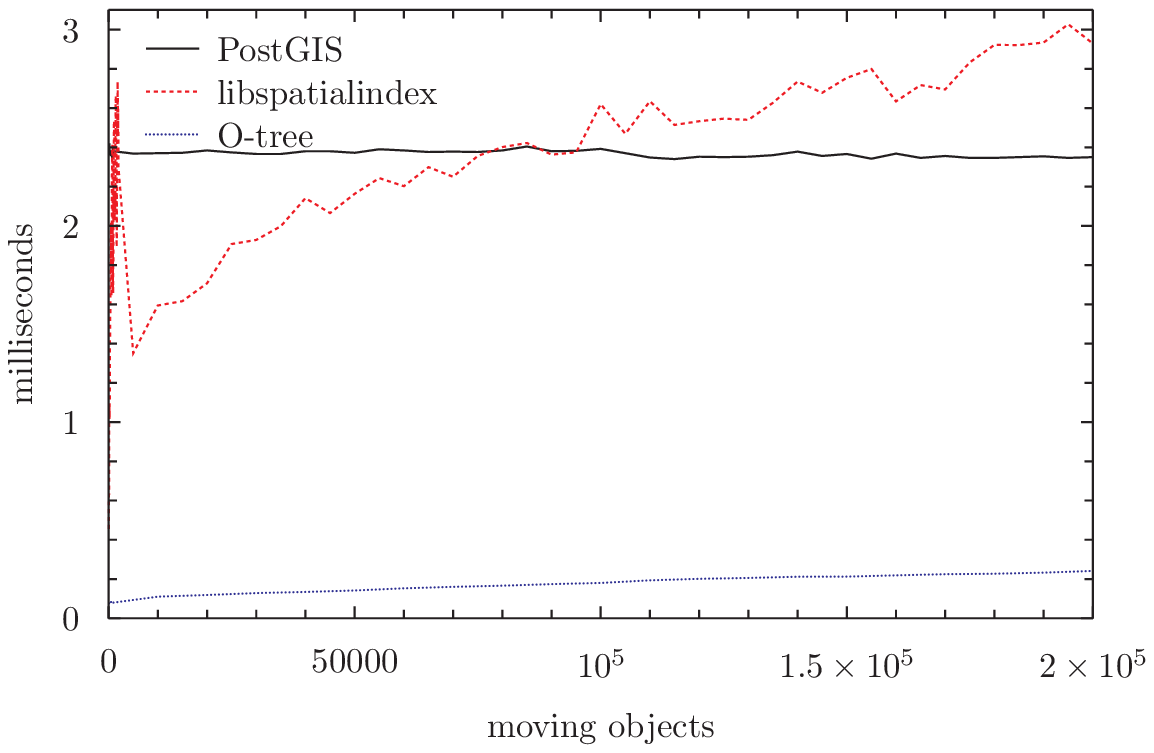}
	} 
	\subfigure[Range Query]{
		\includegraphics[width=0.4\linewidth]{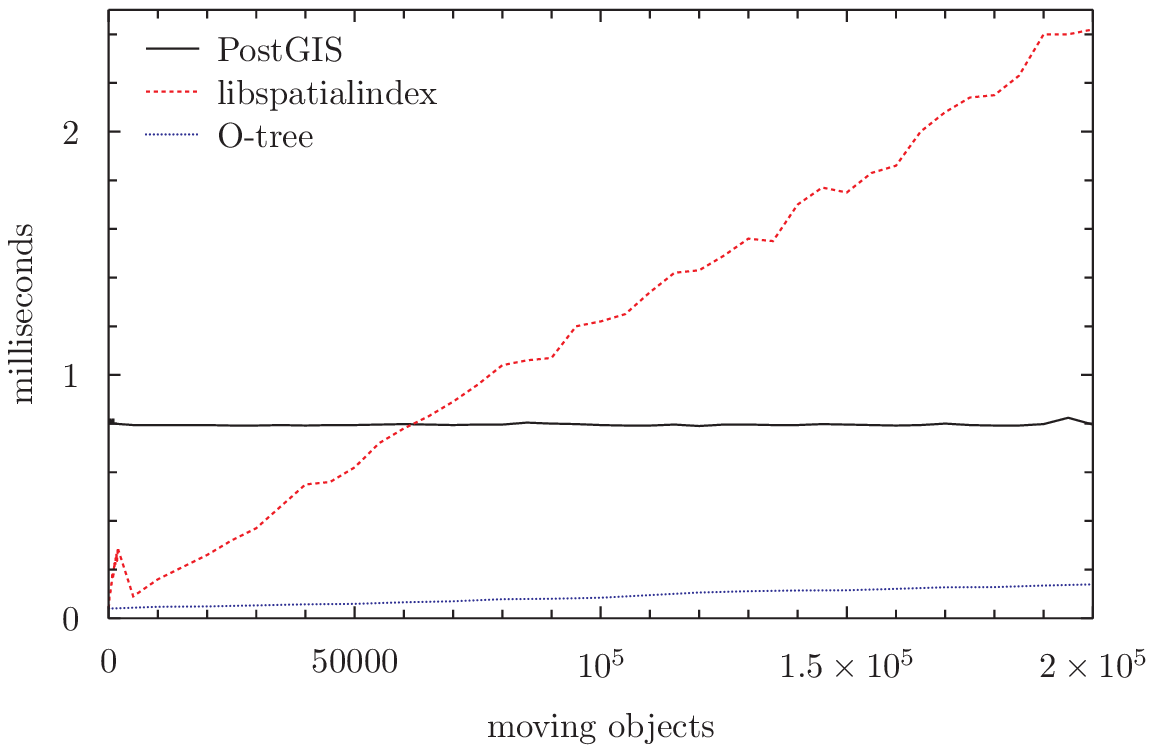}
	}

\caption{R tree implementations} \label{fig:rtrees}
\end{figure}

R-trees are one of the most widely used geometric data structures in geographic information systems (GIS) and spatial databases. 
Their implementations are widespread: From GIS database applications such as MySQL DBMS, PostgreSQL DBMS (spatial extension PostGIS), Microsoft SQL Server, Oracle Spatial, to general purpose libraries such as libspatialindex on unix systems, and publish/subscribe systems. Figure~\ref{fig:rtrees} compares the benchmark results. 

\paragraph{PostGIS}\hspace{-3mm}\cite{postgis} adds support for geographic objects to the PostgreSQL object-relational database. 
As well as most GIS, PostGIS also implements the major geodetic systems, its main target being the geographic applications.

The PostGIS database was entirely stored in the RAM memory.
This is part of the fundamental trade-off and purposes of DSDs; to provide fast answers to spatial queries, essentially dealing with evanescent data, fast trading time but low persistence.
Practically, disc accessing time is far more expensive than the RAM, making a huge difference when dealing with frequent data updates.
Furthermore, in the aforementioned use cases dynamic data are numerous, but they are small, they don't carry much information about the managed objects.
In-memory databases are thus the best storage support for DSDs.
Although widely used for static data processing, PostGIS proves to be too slow when working with a large amount of dynamic data. 

\paragraph{O-tree}\hspace{-3mm}~\cite{Arantes2010,Valero2010} is a Java implementation of R-trees.
It was developed to build several centralized simulators of peer to peer overlays; in particular to showcase the viability of distributed R-tree based publish/subscribe systems.

In a centralized context, each tree is stored in RAM memory.
They provide insert and remove primitives for multidimensionnal objects while supporting range queries expressed with multidimensionnal rectangles.

\subsection{Delaunay 2D triangulation}
\label{sec:delaunay}

A common operation for many geo-spatial applications is the search for nearest neighbors, for example to find the closest 30 people to a given query position. The data is rapidly changing and therefore, it is important to be able to efficiently retrieve proximity information about moving objects.

At the present time most general spatial indexes are data structures based on search trees (R+-tree, R*-tree, m-tree or quadtree). The overall complexity for spatial queries is logarithmic but in the worst case for the kNN is polynomial.

Dealaunay triangulations have been successfully employed in computer vision and graphics for meshing algorithms \cite{DeBerg2008}. Recently it has been used  for triangular P2P network index to process spatial queries \cite{Kang2004}, and for message exchange location data in P2P virtual worlds \cite{Behnoosh}. However, to the best of our knowledge, no general geo-spatial solution for indexing has been proposed. 

In this section we explore the possibility to effectively employ Delaunay triangulations as geo-spatial indexes for moving objects. We claim that they offer a good  performance for all basic operations (insertion, deletion, displacement) and particularly for spatial queries involving nearest neighbor search in an environment with a large amount of moving objects. As support we offer the results for the series of tests described in Section~\ref{sec:experiments}. 

\begin{figure}[!h]
\centering
\includegraphics[width=.5\textwidth]{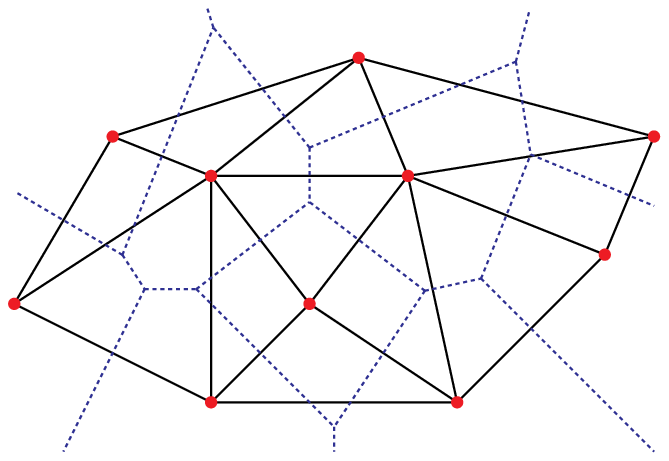}
\caption{Delaunay triangulation and Voronoi diagram}
\label{fig:triangulation}
\end{figure}

A \emph{triangulation} of a set of points $P$ is defined as a maximal planar subdivision whose vertex set is $P$, i.e.,\ no edge connecting two vertices can be added without destroying the planarity.
A \emph{Delaunay triangulation (Delaunay graph)} for $P$ is a triangulation $DT(P)$ such that no point in $P$ is inside the circumcircle of any triangle in $DT(P)$. Moreover, two points form an edge if and only if there is a closed disc that contains the two points on its boundary but does not contain any other point of $P$. Delaunay triangulations enjoy certain properties: they are unique and they maximize the minimum angle of all the angles of the triangles in the triangulation. 

The \emph{Voronoi diagram} is the dual of a Delaunay graph and represents proximity information about the set of objects. The two-dimensional space is partitioned by assigning to each point its nearest object called \emph{generator} or \emph{site}. The points whose nearest sites are not unique will lie on the diagram's edges delimiting the zones of the corresponding object sites. For details 
consult \cite{DeBerg2008}. 


Point location consists in line walking inside the triangulation from an arbitrary object to the query object via the edges. The complexity is $\O{n}$ in the worst case and $\O{\sqrt{n}}$ in average when the objects are distributed uniformly at random.
Any operation on a certain object (e.g.,\ insertion, deletion, displacement) will first locate the vertex inside the triangulation.  Apart from point location procedure, insertion and deletion of one object have linear time complexity with respect to $n$ in the worst case scenario, but constant when the the objects are distributed uniformly at random, see \cite{Devillers2001}.
The nearest neighbor query is similar to a point location, by performing a line walk to the nearest object inside the triangulation. Hence the complexity is $\O{n}$ in the worst case and $\O{\sqrt{n}}$ in average. Computing kNN requires only to check among the neighbors $k$ times. This adds a constant number of operations and does not change the complexity. So, we tested for the benchmark 1NN as spatial query.



\subsection{Triangulation implementations}

\begin{figure}
\centering
	\subfigure[Insertion]{ 
		\includegraphics[width=0.4\linewidth]{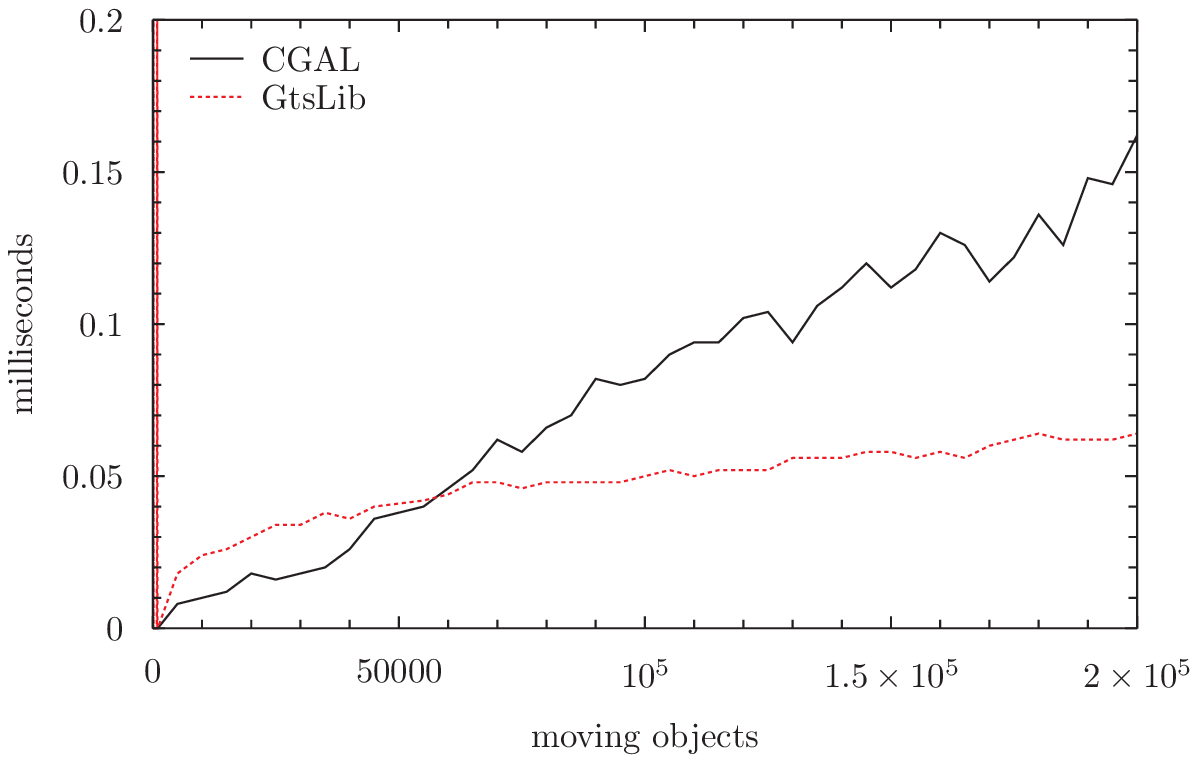}
	}
	\subfigure[Moving]{
		\includegraphics[width=0.4\linewidth]{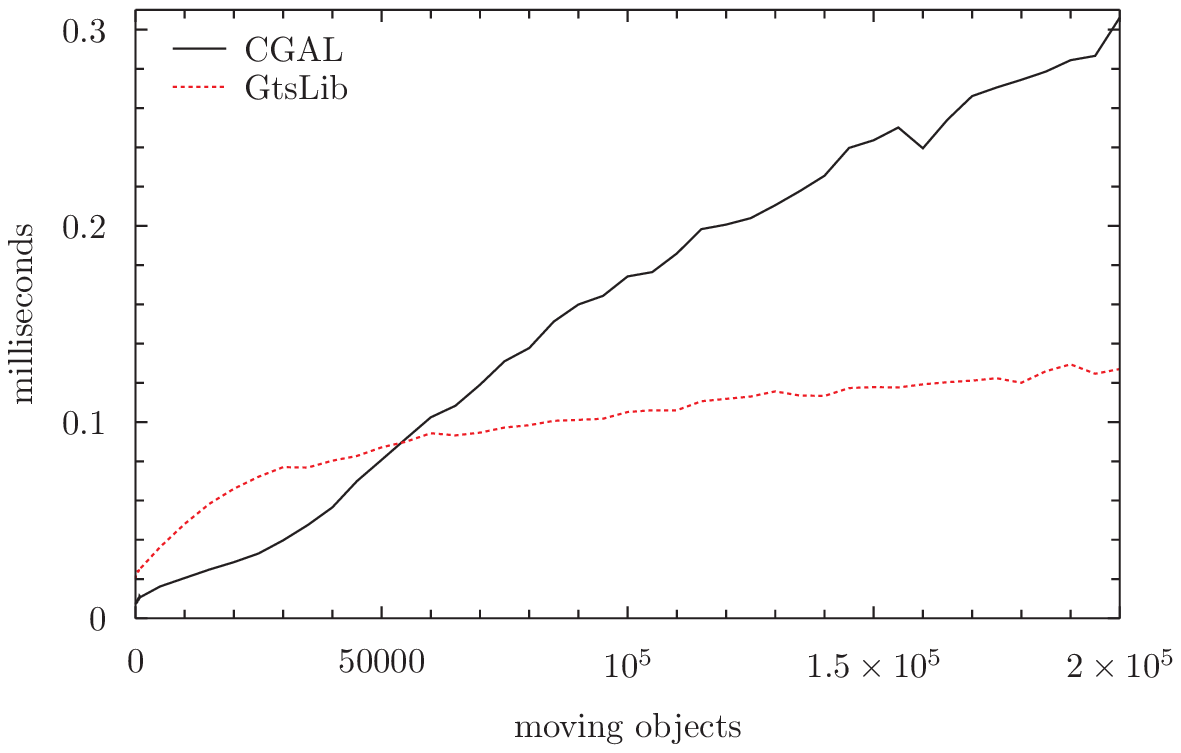}
	} 
	\subfigure[Nearest Neighbor Spatial Query]{
		\includegraphics[width=0.4\linewidth]{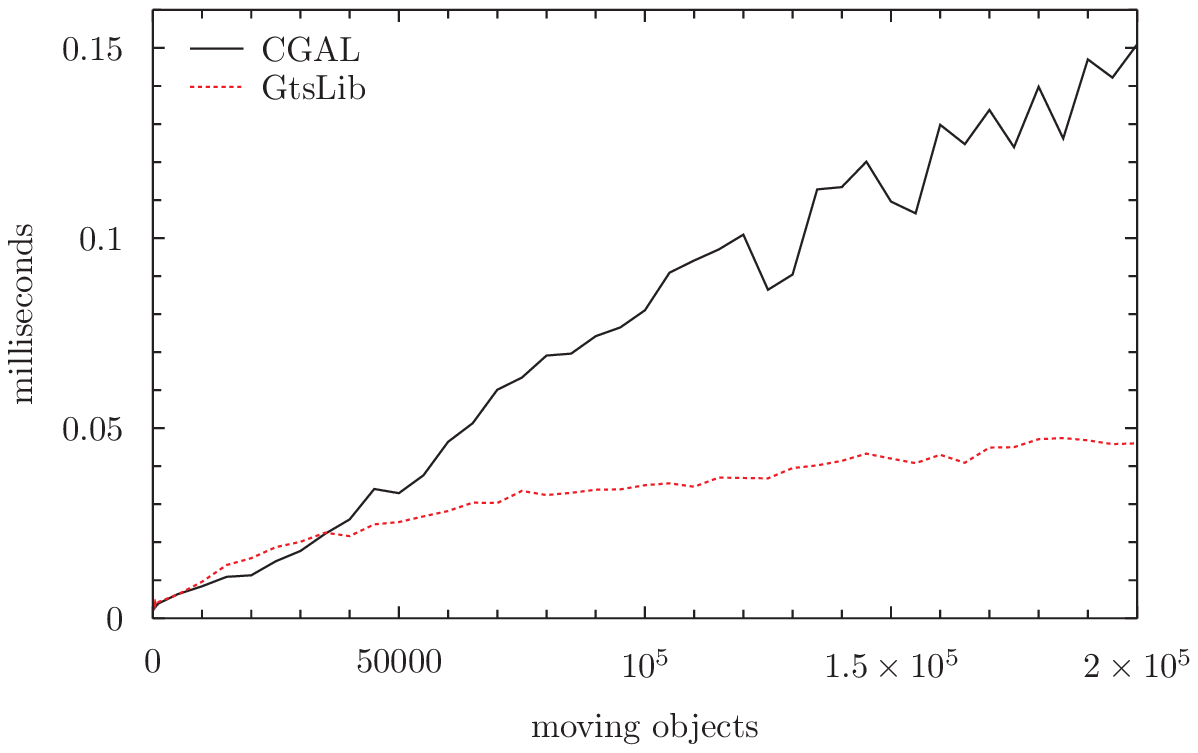}
	}

\caption{The cost for Delaunay triangulation functions} \label{fig:triangulations}
\end{figure}

\paragraph{CGAL}\hspace{-3mm}\cite{cgal} (Computational Geometry Algorithms Library) is a popular scientific tool that offers good performance. It is light and efficiently implemented. The results are described in Figure~\ref{fig:triangulations}.

On the other side, it does not offer the possibility to refine data on attributes other then the spatial ones. Namely, when querying the nearest $k$ neighbors it will provide a quick answer but it will not be able to answer the query of the nearest $k$ neighbors belonging to a certain group of interest.


\paragraph{GTS}\hspace{-3mm}\cite{Library} (GNU Triangulated Surface Library) is a Free Software Library intended to provide a set of useful functions to deal with 2D and 3D surfaces meshed with interconnected triangles.

\subsection{Weaknesses and strengths of both solutions}

The space complexity is in both cases linear with respect to $n$. 
The time complexity is $\O{\sqrt{n}}$ when using Delaunay triangulations and $\O{\ln n}$ for R-trees.

Both solutions enable incremental algorithms which make them suitable for real-time queries on moving objects. Delaunay triangulations have a reduced overhead when a vertex is inserted or removed compared to R-tree algorithms. Moving an object in a triangulation will affect a constant number of nodes and a logarithmic number in an R-tree. Nearest neighbor queries using triangulations are efficient. 

The key characteristic that makes a spatial database a powerful tool is its ability to manipulate spatial data, rather than simply storing and representing it.
The join query enables filtering on different criteria, involving not only spatial ones. Since R-trees are an extension of B-trees join queries can efficiently use an R-tree index as a B-tree index. 
Delaunay triangulation based tools can achieve this at best indirectly, by  employing other tools to refine the spatial result. 

So when querying with other criteria than only position, R-tree --combined with B-trees,-- are, if not mandatory, usually a far better than Delaunay triangulations.

The quintessence of our performance valuation is depicted in Table~\ref{tab:mintime}. We provide a coherent representation of the minimum mean time needed between two successive updates given the total number of mobile objects, summarized for the DSDs described above.

\begin{table}[ht]
\centering
\begin{tabular}{ r || c | c | c | c | c }
 objects &    O-Tree  & spatialidx & PostGIS & CGAL  & GTS \\
  \hline
   50 &  0.004 	&   0.026	&  	   0.12 &  0.001 &  0.001 \\
  500 &  0.04  	&   1   	&	   1.19 &  0.005 &  0.01 \\
   1K &  0.08  	&   2.21  	&	   2.38 &  0.01  &  0.025 \\
   5K &  0.50  	&   6.76  	&	  11.84 &  0.09  &  0.17 \\
  10K &  1.18  	&  16   	&	  25.70 &  0.22  &  0.46 \\
  50K &  8.79  	& 108.15 	&	 118.60 &  4.04  &  4.44 \\
 100K & 21.00  	& 262   	&	 239.20 &  17.22 &  10.70
\end{tabular}
\caption{Minimum time between two updates in seconds}
\label{tab:mintime}
\end{table}

The results of the benchmark illustrate the limits one can achieve with todays systems. These limits are unsatisfactory for certain types of applications described in Section~\ref{sec:applications}.
The existing implementations as well as today's hardware capabilities prove the incapacity to get a satisfactory performance for this kind of demanding real-time applications.
Table~\ref{tab:summarytable} summarizes the scalability limits in terms of number of moving objects for each application scenario.

\begin{table}[ht]
\centering
\begin{tabular}{ r | c || c | c | c | c | c }
  scenario     & T   & O-Tree & spatialidx & PostGIS & CGAL & GTS \\
  \hline\hline\noalign{\smallskip}
          car fleet	&  1min  & 300K & 30K	&   25K   &   650K & 1.5M\\
          	 		&  10s   &  75K	& 7000	&	4000  &    75K & 100K \\
\hline\noalign{\smallskip}
        battlefield	&  1sec  & 8000	& 500	&   400   &    26K & 17K \\
 		awareness	&  10ms  &  150	& 20	&   -     &   900  & 400 \\
\hline\noalign{\smallskip}
              local &  10min &   3M	& 200K	&   250K  &    5M  &  17M\\
  		advertising	&  20s   & 110K	& 12K 	&   8500  &  100K  & 200K \\
\hline\noalign{\smallskip}
             social &  30min &  7M 	& 600K	&   750K  &   70M  & 100M \\
			network	&  20s   & 110K & 12K	&   8500  &  100K  & 200K  \\
\hline\noalign{\smallskip}
             hybrid &  100ms & 2000	& 120	&   30    & 5000   & 2000  \\
     	 reality	&  30ms  &  400	&  60	&   10    & 1000   & 1000  \\
\end{tabular}
\caption{Maximum number of moving objects for each application scenario wehere T is the mean time between two position updates}
\label{tab:summarytable}
\end{table}

Except for today's car fleets, the scalability performances of current dynamic spatial databases are well bellow the needs of the chosen application scenarios in terms of users and update frequency. Furthermore, in many social scenarios filtering on certain parameters is mandatory. This feature makes triangulation based solution difficult to use.

\paragraph{Car Fleets}
This analysis shows that specific software (CGAL and R-tree custom implementations) can handle this amount of data and, with some optimizations and powerful computers, even PostGIS/ramfs can be a solution.  
However, the limits are almost reached and these cannot be acceptable solutions if a better accuracy is needed or if the fleet has more vehicles. Since specific software can be a solution, particular implementations are suitable to handle the amount of data generated by fleets, but no general solution that we know about exists.

\paragraph{Local Advertising}
In marketing scenarios filtering out ad targets by profile is mandatory. This invalidates Delaunay based systems. In the conservative scenario, PostGIS can only take care of few hundreds of thousands of users, far from the millions marketers are used to. With a limit in millions of users, O-Tree could do a better job. However, O-Tree do not currently implement join queries.
Yet DSD's lack of scalability seems to be a major obstacle to the development of location-based advertising.

\paragraph{Location based Social Network, Geo Social Gaming, etc.}
Filtering is mandatory for this activity too. Thus we estimate the maximum amount of simultaneous users that can be handled with current database solutions in thousands to hundreds of thousands, clearly bellow the tens or hundreds of millions of users some of these service have today. Hence, the location-based social on-line arena could greatly benefit from scalable dynamic spatial databases.

\paragraph{Battlefield Awareness}
Actual solutions could handle a reasonable amount of objects at a rough precision. For true performance, a maximum of a few hundreds of objects is considered too small.

\paragraph{Hybrid Reality, Virtual Worlds}
At the rates of 10 to 30 frames per second the systems are mostly useless. Custom systems can handle hundreds of avatars evolving simultaneously. As filtering is not mandatory --usually, partial blindness is considered as a bug-- simple and efficient solutions can be used. And regarding actual virtual world implementations  \cite{Gupta2009} that only scale up to tens of simultaneous avatars per server, the CGAL performance is already good.

However, to build a planetary scale hybrid reality system, with an unbounded number of users, as depicted in \cite{Vinge2006}, this is far from enough and for sure dynamic spatial databases systems need to improve regarding scalability.

\subsection{Summary}
Regarding the scalability, what differs in the above scenarios are the time elapsed between two position updates and the number of simultaneous users. And for most of them, the limits imposed by the existing implementations do not meet the needs. 



\section{Dealing with Scalability Issues}
\label{sec:scalability}
We have seen that for many applications current solutions reach their limits and cannot index the position of all objects at the required frequency. These limitations arise from algorithms that are centralized and rely on one computer --with a fixed amount of processing power, memory, network bandwidth, etc. So, with certainty, for some number $N$ of objects with a given mean frequency $f$ of position updates, the computer will run out of resources.

Recent research works \cite{Mokbel2005,Zimmermann2006,Mouza2007,Gedik2004} have addressed aspects of the scalability of spatial data management but none has brought solutions for soft real-time indexing and querying. For instance \cite{Mouza2007} proposes to split large static datasets to support as many queries as needed.
The replicated parts of the index structure are not timely updated and become outdated when the objects move.
In \cite{Gedik2004} is presented a location monitoring solution in a distributed environment where an important part of the computation is made on the mobile devices. This technique, as well as the peer-to-peer ones \cite{Keller2002, Zimmermann2006}, does not take into account the reduced computational power and energy constraints of mobile devices. 
Moreover, the environment is often heterogeneous: there are many kinds of devices and ways of connecting to the Internet; and taking into account all these differences is costly.

However, it should be possible to scale up to any arbitrary $(N,f)$ combination by adding enough computers to provide the needed resources. In this section we explore approaches for the distribution of load among several machines, but also workarounds to avoid spatial queries on large sets.

\subsection{Cellular approach: Fixed Zones}
The intuitive approach to distribution suggests to divide the problem in smaller problems, each problem being small enough to be handled by a single computer. And in the case of geographic systems it seems natural to partition the territory in contiguous non-overlapping zones, each zone being populated with less moving objects than the maximum manageable by one server. 

Zone servers, each running a dynamic spatial database system, take care of the objects within their zone and process the related spatial queries. Position updates and object inserts are sent only to the zones where they occur. Spatial queries are easily distributed as follows:

\floatname{algorithm}{Distributed Spatial Query for Fixed Zones}
\begin{algorithm}[H]
\renewcommand{\thealgorithm}{}
\caption{}
\begin{algorithmic}[0]
\STATE compute the zone(s) concerned by the query
\FOR { every zone concerned } 
    \STATE run query on the corresponding zone server
    \STATE send back the resulting moving objects
\ENDFOR
\STATE combine the results
\end{algorithmic}
\end{algorithm}

This approach is very efficient when object distribution remains the same over time so each zone is in charge of a roughly constant number of objects. However, many applications start with few objects and then grow up to a stable number. Even assuming that the final configuration is known in advance and stable enough, the fixed zones approach makes the initial phases of deployment and ramp up overly complex leading to unnecessary costs.

But the initial growing phase isn't the only concern: As speaking of moving objects, it is not senseless to expect them to move \emph{en masse} to one place and thus change dramatically the density in a given zone. To take this into account zone servers have to be underloaded to fit these occasional events. As it occurs for cellphone coverage, many zone servers are most of the time merely empty, just waiting for the once-a-week or once-a-month event that will saturate them. 


Besides its lack of flexibility, and depending on the application and the mobility patterns, the costs of overprovisioning
servers can make the fixed zone approach overly costly and unusable.

\subsection{Partial solutions and workarounds}
For particular applications partial solutions have emerged. These solutions are different ways of dealing with the fact that dynamic spatial database systems are not scalable and they do not actually intend to solve the DSD scalability issue.

\paragraph{Sharding/instancing in virtual worlds} Actual solutions for virtual worlds can only host at most a few hundreds simultaneous users on a single server \cite{Gupta2009}. To accommodate more users the usual solution is to run more ``shards" or servers, each one hosting an ``instance" of the virtual world. Merely, this workaround consist in having multiple copies of the virtual world, each copy being limited in the number of concurrent users interacting with each other. Moreover, as users on one instance cannot interact with users on another instance, this workaround is only useful for games or services that can integrate these constraints in their scenarios.

\paragraph{Named location zones} Other services avoid spatial queries altogether. They use instead, the notion of named location zones, also called \emph{venues} or \emph{check-ins}. In this schema users register to a venue and, to find who is nearby, they query who is registered in the same or neighboring venues.
This assumes the granularity of the named zones to match the density of users, if not the system may return too many or too few neighbors, making the answer irrelevant.
This is a heavy constraint on the named zones that need also to be meaningful for the users and to exist all over the territory the service is intended to operate. The design of named zones is an incredibly difficult task, poised to give imperfect results.

\paragraph{Limited list} Another common workaround to is to make the spatial query on a reduced set of moving objects. Typically, this is the case with the services that propose to find which ``friends" are nearby. Since the number friends rarely exceeds hundreds, checking among this number who is nearby is easily feasible.

\section{Scalable Dynamic Spatial Database System}
As we have seen, the limits in scalability of dynamic spatial databases take root in the computational costs of continuously spatially indexing the many moving objects. 
Dividing the problem into smaller problems, each one involving a number of moving objects small enough to be handled by a single machine, will allow the indexing process to scale.

\subsection{Self Adaptive Zones}

That's exactly what the fixed zone approach intend to do. But since fixed zones can only be related with statistical properties (position, movement patterns,...) of the moving objects, they fail to capture dynamic events. This may lead to situations where zones are almost empty or otherwise overloaded.

However, spatially dividing the problem in non-overlapping zones still makes sense for spatial queries.
A characteristic feature of most spatial queries, like the $k$ nearest neighbor(s) or range search queries, is to span over a limited portion of the space. Therefore, a given spatial query will apply to a limited subset of zones servers only.

In our solution --named Scalable Dynamic Spatial Database or SDSD-- we have combined both properties: each server is responsible for indexing the positions of a set of moving objects and each server indexes all moving objects of a contiguous zone. As objects move, leave or join the system, the set of objects --as well as the covered zone-- changes in order to balance the load and maintain the contiguity of the zone.

\floatname{algorithm}{Zone Resizing / Load Balancing}
\begin{algorithm}[H]
\renewcommand{\thealgorithm}{}
\caption{}
\begin{algorithmic}[0]
\STATE zone servers send their load levels to the neighboring zones
\STATE some moving objects might change their belonging zones
\STATE \hspace*{1cm} to balance load between zone servers
\STATE \hspace*{1cm} to maintain contiguity and compactness of zones
\end{algorithmic}
\end{algorithm}

Each zone server implements a DSD and maintains a spatial index for its subset of moving objects; hence it is able to process spatial queries on this subset.

\subsection{Spatial Query Dispatcher}

To process a spatial query, a special module, the spatial query dispatcher, selects zone servers to forward the incoming queries. The spatial query dispatcher collects the answers from the zone servers and combines them to build the final answer.

The trivial way of computing the zones concerned by a spatial query is to have a map of all the zones. However, this solution is impractical since the shape of a zone changes frequently and can be a complex polygon with up to $n$ vertices, where $n$ is the size of the indexed subset. Instead, we have chosen to represent each zone by a sample object belonging to the zone.

\tikzstyle{server} = [rectangle, draw, text width=5em, text centered,  minimum height=4em]
\tikzstyle{shadow_server} = [server, node distance = 1mm]

\tikzstyle{line} = [draw, -latex']
\tikzstyle{mo} = [draw, ellipse,text width=3.4em, minimum height=5em]
\tikzstyle{shadow_mo} = [mo, node distance = 0.75mm]

   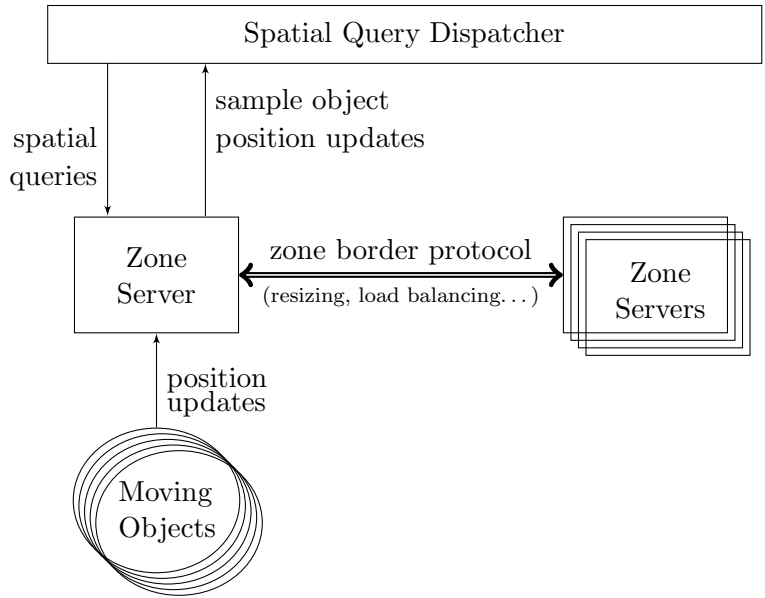
\begin{figure}[h]
   \centering
\begin{tikzpicture}[ auto]
    \node [server] (zs1) {Zone Server};
    \node [server,right of=zs1, node distance = 6.5cm] (zs2) {};
    \node [shadow_server,right of=zs2, below of=zs2] (zs3) {};
    \node [shadow_server,right of=zs3, below of=zs3] (zs4) {Zone Servers};
    \node [shadow_server,right of=zs4, below of=zs4] (zs5) {};
   \path [double, line,<->,thick] (zs1) -- node{zone border protocol}node[below,font=\scriptsize]{(resizing, load balancing\dots)}(zs2);
   
   \node[mo, below of=zs1,  node distance = 3cm](mo1){};
    \node [shadow_mo,right of=mo1, below of=mo1] (mo2) {};
    \node [shadow_mo,right of=mo2, below of=mo2] (mo3) {Moving Objects};
    \node [shadow_mo,right of=mo3, below of=mo3] (mo4) {};
    \node [shadow_mo,right of=mo4, below of=mo4] (mo5) {};
  
   \path [line] (mo1) -- node[right]{position}node[right][near start]{updates}(zs1);

\node[ node distance = 0.65cm, above of=zs1,right of=zs1](zs11){};
\node[ node distance = 0.65cm, above of=zs1,left of=zs1](zs12){};
\node[ node distance = 2.3cm, above of=zs11](r1){};
\node[ node distance = 2.3cm, above of=zs12](r2){};
   \path [line] (zs11) edge
   		node[right][near end]{sample object}
   		node[right]{position updates}
  		node[right][near start]{}(r1);
\path [line] (r2) edge node[left]{spatial}node[left][near end]{queries}(zs12);

\node[node distance = 3.2cm,draw,rectangle, above of=zs1,left of=zs2, text centered,text width=24em, minimum height=2em](qd){Spatial Query Dispatcher};

\end{tikzpicture}
   \caption{Scalable Dynamic Spatial Database Architecture}
   \label{fig:Architecture}
   \end{figure}

Each zone server sends the updates of its sample object position when it moves, or promotes a different one when needed. Hence, the spatial query dispatcher, which implements a DSD, maintains a up-to-date spatial index for the set of sample objects.

\floatname{algorithm}{Spatial Query Dispatch}
\begin{algorithm}[H]
\renewcommand{\thealgorithm}{}
\caption{}
\begin{algorithmic}[0]
\STATE $S:$ set of sample moving objects
\STATE \textit{(each zone is represented by one sample object)}
\STATE run the spatial query on $S$
\STATE \hspace*{0.8cm} $\rightarrow$  select zones
\FOR { every selected zone } 
    \STATE run query on the corresponding zone server
    \STATE \textbf{if} query intersects the border \textbf{then}
    \STATE \hspace*{1.2cm} forward query to the neighboring zone
    \STATE send back the resulting moving objects
\ENDFOR
\STATE combine the results
\end{algorithmic}
\end{algorithm}

When it receives a spatial query --nearest neighbor(s) or range searching-- the dispatcher finds which moving object(s) of the sample set match the query. And a zone corresponding to a matching sample object is concerned by the query: for a nearest neighbor search query because the matching sample object is a first approximation so the final result is nearby and hopefully in the same zone; and for a range search query because as one object of the zone is in the range it's a good expectation to find more in the same zone. 
Then the query is issued to the matched zones and, occasionally, within a zone server, if a query reaches a border it is forwarded to the neighboring zone.

\subsection{Dynamic Zones}
In our system, zone servers are responsible for the spatial indexing of a set of moving objects. Also, each server defines a contiguous zone, i.e., a zone in one piece that changes dynamically. The relation between the zone and the set of indexed objects is the following: every object in the zone is indexed by the server and every indexed object is in the zone.

To avoid superfluous redundancy, objects belong to one zone server only. This means zones must not overlap. Hence, as objects move, join or leave the system, zone servers continuously exchange information to maintain zones that do not overlap.

When using a DSD based on a Delaunay triangulation we have come up with an elegant solution for defining zone borders: the zone is defined as the union of the Voronoi cells of the indexed objects. As the Voronoi diagram is the dual of the Delaunay triangulation used for the spatial index, the zone borders are recomputed whenever the object set changes.

For our particular solution based on Delaunay triangulations, we are implementing the details of the zone border protocol and how zone servers decide which objects are indexed by which server and we plan to release and evaluate this first implementation soon.

\subsection{A Generic Scalable Architecture}
As shown in Figure~\ref{fig:Architecture}, this architecture for scalable dynamic spatial database systems with a spatial query dispatcher and many zone servers represented by sample objects is generic. The policies for load balancing and zone shaping/resizing may vary to adapt to particular applications. As R-trees and Delaunay triangulations have their own strengths and drawbacks, the underlying dynamic spatial database systems used for indexing the moving objects in the zone may also change.

Arguably, the dispatcher is a bottleneck that could impede the system to scale further. However, if the actual limit for a given update frequency is $n$ moving object per DSD, the dispatcher can index $n$ sample objects and therefore manage $n$ zone servers each one indexing $n$ moving objects. So, the limit for a single level dispatcher is $n^2$ moving objects. This could quite enough for many applications. But if needed, the DSD within the dispatcher could be as well be implemented using zone servers and an upper level spatial query dispatcher and so forth.

A $k$-level hierarchical SDSD will hence scale up to $n^k$ moving objects with a latency in $\O{k}$ on queries. To put it more vivid, the hybrid/virtual reality scenario of table \ref{tab:summarytable} will scale up to one billion simultaneous avatars/users in only 3 levels.

\section{Concluding Remarks and Future Work}
In this paper we study the existing software solutions for dynamic spatial databases --i.e.,\  systems intended to index moving objects positions and answer spatial queries. The measurement of performance of actual systems leads to the following conclusion: regarding the massive amount of position data available today, actual systems face a huge scalability problem.

Solving these scalability issues resides in distributing the load among enough servers. We have designed a distributed architecture for scalable dynamic spatial databases and the next steps are implementing and testing these algorithms.

In parallel, as very recent works \cite{Maljovec2010, Rong2008} aim at implementing the computation of 2D Delaunay graphs using GPUs\footnote{Graphical Processing Unit}.
Since GPU can be dramatically faster than CPU for some geometrical problems, it should be interesting to evaluate this implementation when available. With GPU based implementations of DSDs, the per moving object costs could be lowered by one order of magnitude making our solution even less costly.

We hope in a near future new services, as a planet wide hybrid reality, will be possible to implement thanks to scalable DSDs.


\bibliographystyle{abbrv}
\bibliography{1nn}
\end{document}